\documentstyle[twoside,fleqn,espcrc2,epsfig]{article}

\makeatletter
\def\setcaption#1{\def\@captype{#1}}
\makeatother
\newcommand{\AmS}{{\protect\the\textfont2
  A\kern-.1667em\lower.5ex\hbox{M}\kern-.125emS}}

\title{A new phase structure in three-dimensional dynamical triangulation model}

\author{T.~Hotta\address{Institute of Physics, University of Tokyo, Meguro-ku,
Tokyo 153, Japan},
\underline{T.~Izubuchi\address{Institute of Physics, University of Tsukuba, Tsukuba, Ibaraki 305, Japan}} and
J.~Nishimura\address{Department of Physics, Nagoya University, Chikusa-ku, Nagoya 464-01, Japan}
}       
\begin{document}
\begin{abstract}
We investigate a new phase structure 
of the three-dimensional dynamical triangulation model
with an additional local term by Monte Carlo simulation.
We find that the first order phase transition observed
for the naive Einstein-Hilbert action
terminates at a finite coefficient of the new term.
The phase transition turns into second order at the endpoint, beyond
which it becomes a crossover.
\end{abstract}

\maketitle

\section{Introduction}

The success of the dynamical triangulation model in two dimensions
summarized by its equivalence to the Liouville theory in the continuum limit
motivated intensive studies of the model in higher dimensions
through Monte Carlo simulation.
It turned out, however, that 
with an action obtained by naive discretization of the Einstein-Hilbert 
action, the phase transition observed is of first order
both in three dimensions and in four dimensions \cite{4D1stPT}.
In lattice theories in general, it is often preferable 
(\em e.g. \em improved actions to reduce the finite lattice spacing effects)
or even necessary (\em e.g. \em the Wilson term to eliminate doublers)
to modify the lattice actions from that obtained by naive discretization
of the action in the continuum.
We may therefore hope to obtain a second order phase transition
in the present case
by adding a local term to the naive Einstein-Hilbert action.

The new term we consider is motivated by a possible contribution
from the measure for the path integral over the metric
and has been first studied in four dimensions \cite{BrugmannMarinari}.
Recently this modified action has been studied 
in three dimensions by the Monte Carlo renormalization group
and a possibility was suggested that
the first order phase transition terminates at a finite coefficient
of the new term \cite{Renken}
\footnote{Note that our $\mu$ is the minus of the 
$\mu$ in Ref. \cite{Renken}}.
We clarify this possibility by high statistics Monte Carlo simulation
using multi-canonical technique.

\section{The model}

The naive discretization of the Einstein-Hilbert action
in the dynamical triangulation model in \( D \) dimensions 
gives
\( S_{0}=-\kappa _{0}N_{0}+\kappa _{D}N_{D} ,\)
where \( N_{0} \) and \( N_{D} \) are the
number of vertices and \( D- \)simplices respectively 
in the simplicial manifold.
We modify the action as
\begin{eqnarray}
\label{eq:totAction} S & = & S_{0}+\mu M\, ,\\
\label{eq:MUterm} M & = & \sum _{v}\log \left[ o(v)/(D+1)\right] \, ,
\end{eqnarray}
 where \( \mu  \) is the new coupling constant, and 
\( o(v) \) is the number of \( D- \)simplices
that contain the node \( v \). 
The summation in (\ref{eq:MUterm}) is taken over all the nodes in the
manifold. 
This term \cite{BrugmannMarinari} 
can be considered as the contribution from the path integral 
measure \( \Pi _{x}\, g(x)^{-\mu /2} \),
since \( o(v) \) can be interpreted as the ``local volume'',
which corresponds to \( \sqrt{g(x)} \) in the continuum.

In the following we consider the canonical ensemble with
fixed $N_D$ given by the partition function :
\begin{eqnarray}
\label{eq:PartFunc1}
Z(\kappa _{0},\mu ) &=&  \sum _{\{T(N_{D})\}}
\exp (\kappa_0 N_0 -\mu M) \, ,\\
\label{eq:PartFunc} &=& 
\sum _{N_{0}}z(N_{0};\mu )\, \exp (\kappa _{0}N_{0})\: ,
\end{eqnarray}
where the summation in (\ref{eq:PartFunc1})
is taken over all possible triangulations of 
\( D \)-dimensional sphere
that contain \( N_{D} \) \( D \)-simplices. 
The $z(N_{0};\mu)$ is the partition function for fixed $N_0$,
which can be given by
\begin{eqnarray}
\label{eq:SpectFunc} z(N_{0};\mu)  
=  \sum _{\{T(N_{D},N_{0})\}}
\exp (-\mu M)\: ,
\end{eqnarray}
where the summation is now taken over
all possible triangulations 
that contain \( N_{D} \) 
\( D \)-simplices and $N_0$ vertices.
The details of the numerical simulation will be reported elsewhere
\cite{forthcoming}.

\section{End point of the first order phase transition}
\begin{center}
{\centering \epsfig{file=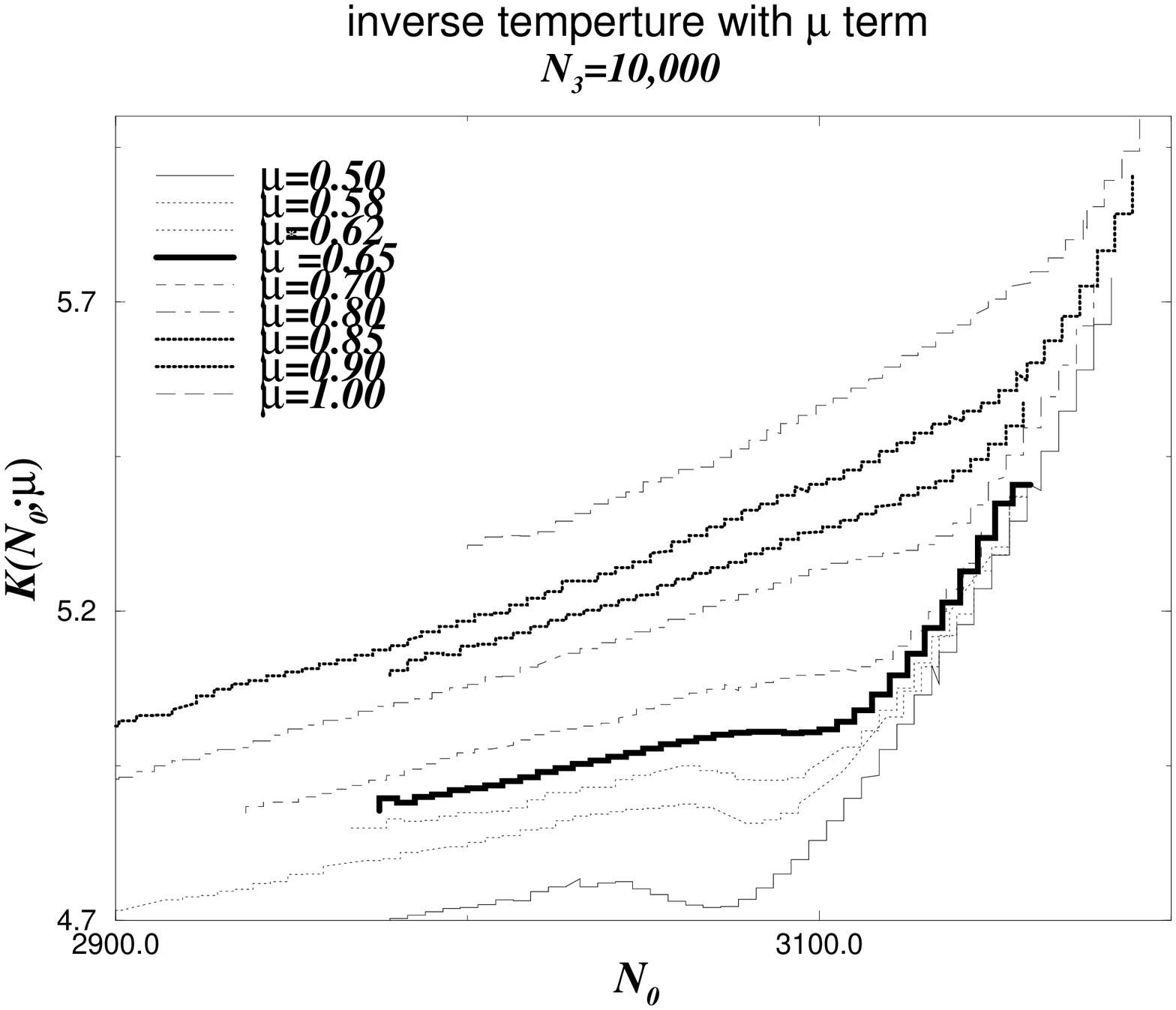, width=0.4\textwidth} }
\setcaption{figure}
\caption{The microcanonical inverse temperature 
\protect\( K(N_{0};\mu )\protect \) 
as a function of \protect\( N_{0}\protect \) for several 
\protect\( \mu \protect \)'s
with the system size \protect\( N_{3}= 10,000\protect \).
\label{fig:K}}
\end{center} 

In order to investigate the phase diagram
with the modified action,
we measure the microcanonical entropy 
\( s(N_{0};\mu ) \)
defined by
\begin{equation}
\label{eq:EnthropyDef} 
s(N_{0};\mu )  =  \log  z(N_{0};\mu ),
\end{equation}
from which we calculate the 
microcanonical inverse temperature \( K(N_{0};\mu ) \) by
\begin{equation}
\label{eq:KappaDef} K(N_{0};\mu ) 
 =  s(N_{0}+1;\mu )-  s(N_{0};\mu ).
\end{equation}
Precise measurement of the microcanonical entropy
near the first order phase transition point is possible
by using the multi-canonical technique.
This was previously done
for the unmodified action with $\mu=0$ in Ref. \cite{Fujitsu},
and the critical exponents consistent with
the first order phase transition
have been successfully extracted by the finite-size scaling analysis.

In Fig. \ref{fig:K}, we show the microcanonical inverse temperature
\( K(N_{0};\mu ) \) for several values of \( \mu  \).
Note that the 
node distribution is given for each $\kappa_0$ and $\mu$
by $ z(N_{0};\mu ) \exp (\kappa _{0}N_{0})$
and the double peak structure in the
node distribution,
which is a signal of the first order phase transition, 
can be probed directly by looking at the microcanonical inverse temperature
\( K(N_{0};\mu ) \). The double peaks appear if and only if 
(i) \( K(N_{0};\mu ) \) has both local maximum 
and local minimum and 
(ii) \( \kappa _{0} \) is set to a value between
the local minimum and the local maximum. 
One can see that the local minimum and the local maximum observed
for $\mu=0$ \cite{Fujitsu} merge 
at \( \mu ^{*}=0.65 \),
beyond which the \( K(N_{0};\mu ) \) becomes a monotonous function of
$N_0$.
This suggests that the first order phase transition ceases to exist
at \( \mu ^{*} \), beyond which it becomes a crossover.
\begin{center}
{\centering \epsfig{file=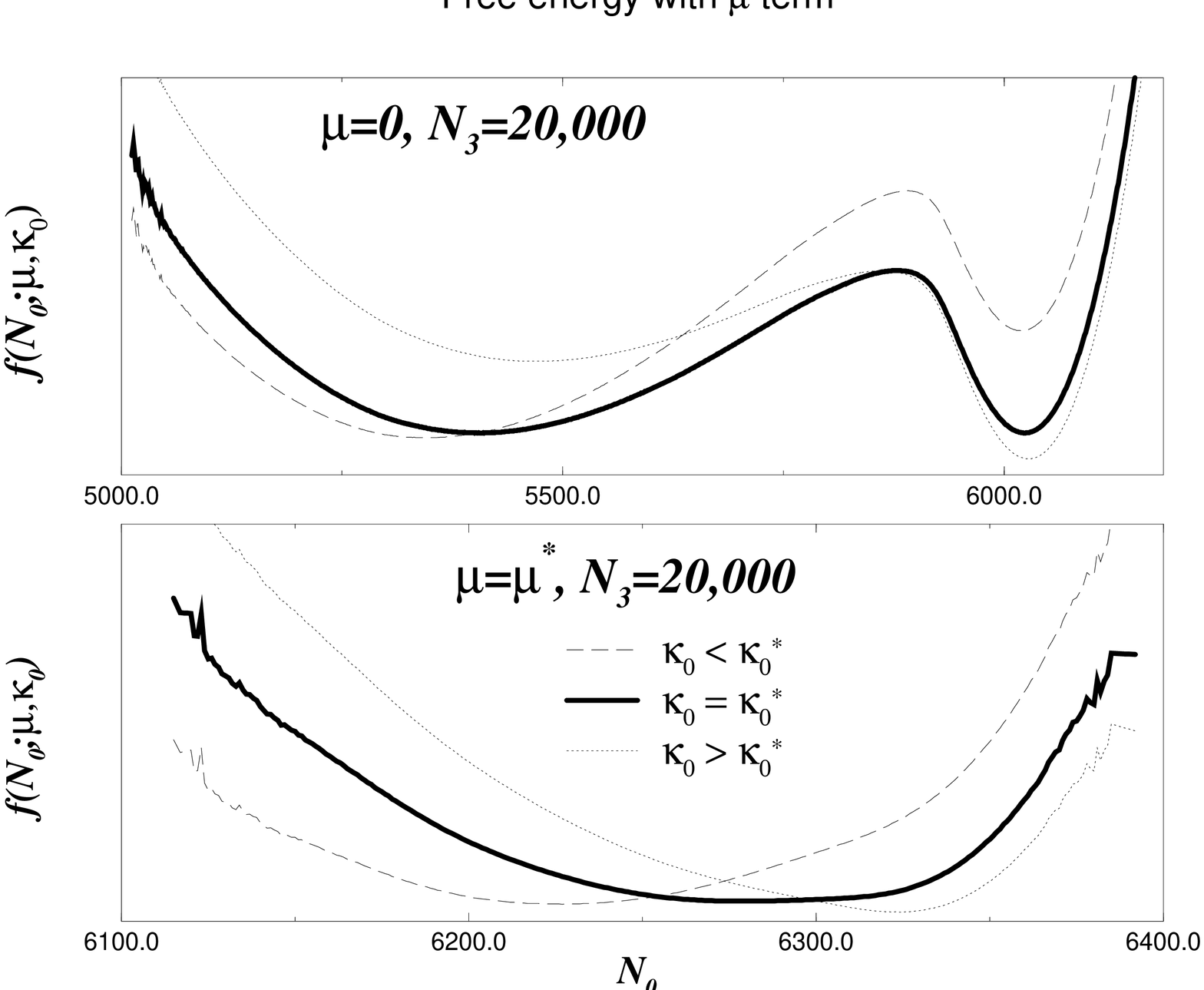, width=0.4\textwidth}}
\setcaption{figure}
\caption{The free energy \protect\( f(N_{0};\, \mu ,\kappa _{0})\protect \) 
as a function of $N_0$
near $\kappa_0^{*}$ for $\mu=0$ and $\mu=\mu^{*}$
with the system size $N_3=20,000$.
\label{fig:FreeEnergy}}
\end{center}

For each \( \mu  \) below \( \mu ^{*} \),
we define the critical coupling \( \kappa _{0}^{*} \)
of the phase transition
as the position of the peak of the node susceptibility 
\(  \chi (\kappa _{0},\mu ) = 
\left\langle N^{2}_{0}\right\rangle 
-\left\langle N_{0}\right\rangle ^{2} \).
The end point of the phase transition line is given by 
\( \mu ^{*}=0.65,\, \kappa ^{*}_{0}=5.00 \) 
for \( N_{3}=10,000 \), and 
\( \mu ^{*}=0.80,\, \kappa ^{*}_{0}=5.515 \) for \( N_{3}=20,000 \).
In Fig. \ref{fig:FreeEnergy}, we plot the free energy
\( f(N_{0};\, \mu ;\kappa_0) = -s(N_{0};\mu )-\kappa _{0}N_{0} \),
which is nothing but the minus log of the node distribution,
as a function of $N_0$ near $\kappa_0^{*}$ for $\mu=0$ and $\mu=\mu^*$.
The two local minima observed for \( \mu =0 \) merge into one at
\( \mu =\mu ^{*} \).
The free energy curve at 
\( (\kappa ^{*}_{0},\mu ^{*}) \) has a wide flat
bottom, which suggests the existence of a massless mode.


\ \vspace*{-0.3cm}
\section{Fractal structure at the end point}
In order to examine a possible continuum limit at the end point,
we measure the boundary area distribution
\( \rho (A,r) \), which is the number of boundaries 
with the area \( A \) at the geodesic distance \( r \).
The corresponding quantity in two dimensions
called the loop length distribution
is calculated analytically and is found to
possess a continuum limit \cite{KKMW}.
The scaling behavior expected from this result
has been correctly reproduced by numerical simulation
\cite{TsudaYukawa}.

In Fig. \ref{fig:BAD}, we plot the boundary area distribution
at the end point of the first order phase transition line
as a function of \( x=A/r^{2} \). 
\begin{center}
{\centering \epsfig{file=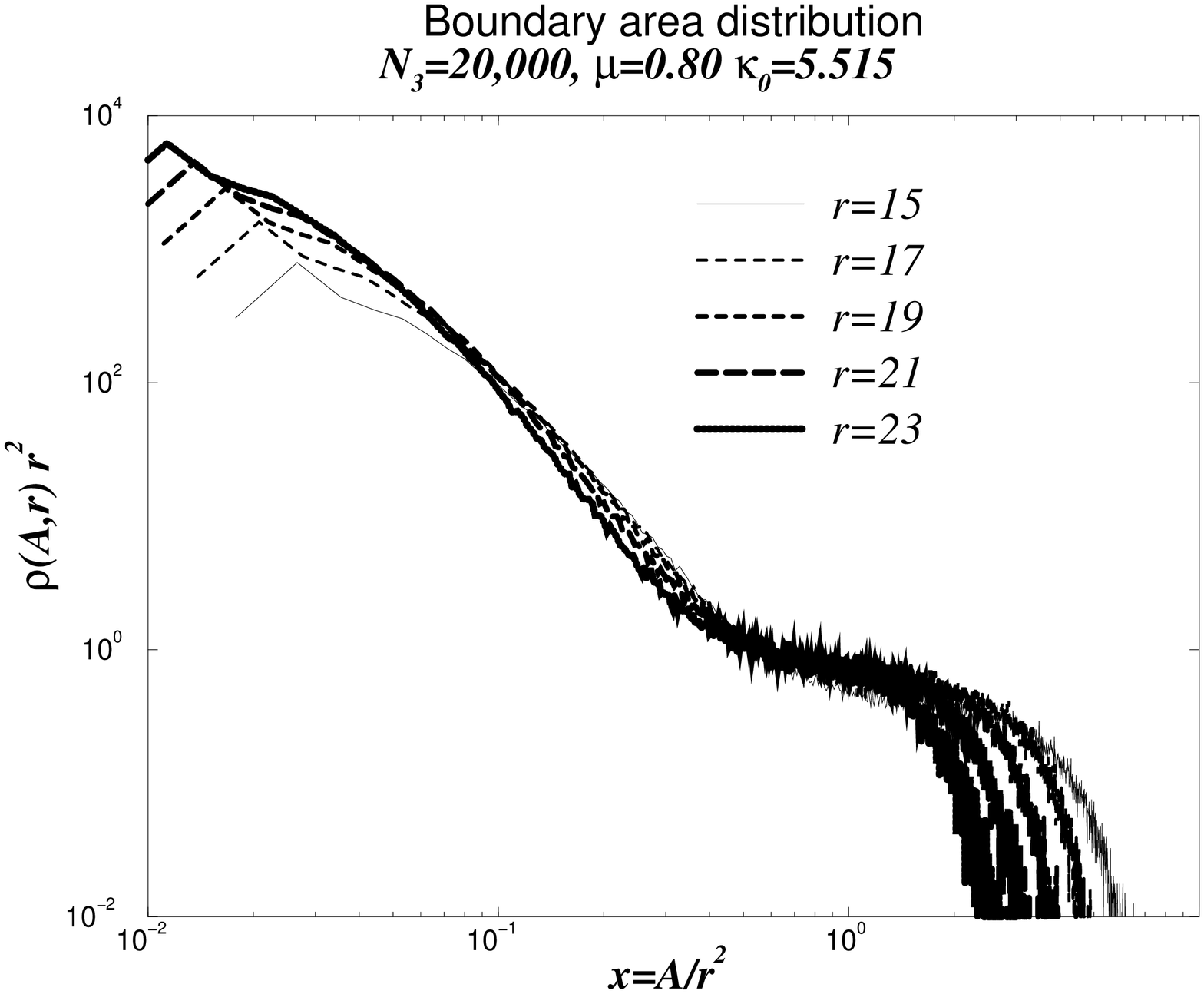, width=0.4\textwidth} }
\setcaption{figure}
\caption{Boundary area distribution \protect\( \rho (A,r)\protect \) 
at the end point of the first order phase transition line
with the system size $N_3=20,000$.
\label{fig:BAD}}
\end{center}
One can see a reasonable scaling behavior.
The power of $r$ in the scaling variable \( x=A/r^{2} \)
implies that the fractal dimension is 3.
We would like to remark that the scaling behavior we observe
is much better than the one that has been claimed to exist with
the unmodified action \cite{Hagura}. 

\ \vspace*{-0.3cm}
\section{Discussions}

Our results suggest that the first order phase transition
becomes second order at the end point,
where one can obtain a continuous theory.
A natural question to be asked is what the continuum theory is.
One may suspect that we cannot construct any continuum theory
in three-dimensional quantum gravity, since 
we have no physical degrees of freedom.
This argument is too naive, however.
An example of field theory which has no physical degrees of freedom
that is still completely well-defined 
is two-dimensional Yang-Mills theory.

The fact that we have to fine-tune two parameters to obtain the 
continuum limit implies that there are two relevant operators
around the fixed point.
A possible interpretation of the continuum theory is therefore 
the \( R^{2} \) gravity. 
The observation that the fractal dimension at the fixed point
is approximately three is also suggestive of this interpretation.
The fractal dimension we extracted, however, is still preliminary,
and the value might increase as we increase the system size.
Needless to say, a lot of work is yet to be done to
understand the nature of the continuum theory.

In 4D, the situation is different from the 3D case in that
singular vertices which are shared by huge number of $D$-simplices
exist in the strong coupling phase \cite{singular}.
We can add a local term in the action that eliminates these singular vertices,
if one wishes, and then the situation becomes more or less the same 
\cite{forthcoming}.
We therefore expect that the first order phase transition can be changed into
second order by modifying the action.

\end{document}